\documentclass[prb,twocolumn,amsmath,amssymb,floatfix,footinbib,bibnotes]{revtex4-1}

\pdfoutput=1

\usepackage[colorlinks=true,citecolor=blue,linkcolor=magenta]{hyperref}
\usepackage{amsmath}
\usepackage{graphicx}
\usepackage{dsfont}
\usepackage{changepage}
\usepackage{appendix}
\usepackage[T1]{fontenc}
\usepackage[latin1]{inputenc}
\usepackage{lipsum}

\newcommand{\be}{\begin{equation}}
\newcommand{\ee}{\end{equation}}

\def \ua{{\uparrow}}
\def \da{{\downarrow}}
\def \be{\begin{equation}}
\def \ee{\end{equation}}
\def \ba{\begin{array}}
\def \ea{\end{array}}
\def \bea{\begin{eqnarray}}
\def \eea{\end{eqnarray}}
\def \nn{\nonumber}

\def \e{{\varepsilon}}
\def \lam{{\lambda}}
\def \L{{\Lambda}}
\def \a{{\alpha}}

\def \g{{\gamma}}
\def \D{{\Delta}}
\def \d{{\delta}}
\def \w{{\omega}}
\def \s{{\sigma}}
\def \f{{\varphi}}

\def \e{{\varepsilon}}

\def \G{{\Gamma}}

\def \beas{\begin{eqnarray*}}
\def \eeas{\end{eqnarray*}}

\def \mrm{\mathrm}

\def \bs{\boldsymbol}
\def \mc{\mathcal}

\newcounter{indice}

\begin{document}
\title{Ferromagnetic and Nematic Non-Fermi Liquids in Spin-Orbit Coupled Two-Dimensional Fermi Gases}
\author{Jonathan Ruhman and Erez Berg \\
{\small \em Department of Condensed Matter Physics, Weizmann Institute of Science, Rehovot 76100, Israel}}
\begin{abstract}
We study the fate of a two-dimensional system of interacting fermions with Rashba spin-orbit coupling in the dilute limit. The interactions are strongly renormalized at low densities, and give rise to various fermionic liquid crystalline phases, including a spin-density wave, an in-plane ferromagnet, and a non-magnetic nematic phase, even in the weak coupling limit. The nature of the ground state in the low-density limit depends on the range of the interactions: for short range interactions it is the ferromagnet, while for dipolar interactions the nematic phase is favored. Interestingly, the ferromagnetic and nematic phases exhibit strong deviations from Fermi liquid theory, due to the scattering of the Fermionic quasi-particles off long-wavelength collective modes. Thus, we argue that a system of interacting fermions with Rashba dispersion is generically a non-Fermi liquid at low densities.
\end{abstract}
\maketitle
\section{ Introduction  }
The realization of strongly spin-orbit coupled fermion systems in low dimensions, either in solid state or cold atomic setups,\cite{Ast2007,Caviglia2010a,Wray2010,Wang2012,Cheuk2012,Campbell2011} calls for an understanding of the interplay between many-body interactions and spin-orbit coupling. One of the effects of spin-orbit coupling in solids is to modify the dispersion relation of electrons; as a result, inter-particle interactions can be effectively enhanced. For example, in the case of Rashba-type spin orbit coupling (which occurs in two-dimensional electron gases in quantum wells without inversion symmetry), the dispersion minimum occurs on a nearly-degenerate \emph{ring} in momentum space, instead of a single minimum at $\bs{k}=0$. This leads to quenching of the kinetic energy at low densities, and hence many-body interactions become increasingly important. It has been argued that in the low-density limit and in the presence of short-range repulsive interactions, a host of ``electronic liquid crystal'' phases can be stabilized,\cite{Berg2012} including nematic, ferromagnetic nematic, and anisotropic Wigner crystal phases.~\cite{Berg2012,Silvestrov2014} In bosonic systems, Rashba spin-orbit coupling can lead to unusual phases, as well.~\cite{Wang2010,Wu2011,Gopalakrishnan2011,Jian2011,Barnett2012,Sedrakyan2012,Zhou2013}

\begin{figure}
\centering
\includegraphics[width=8.5cm,height=8cm]{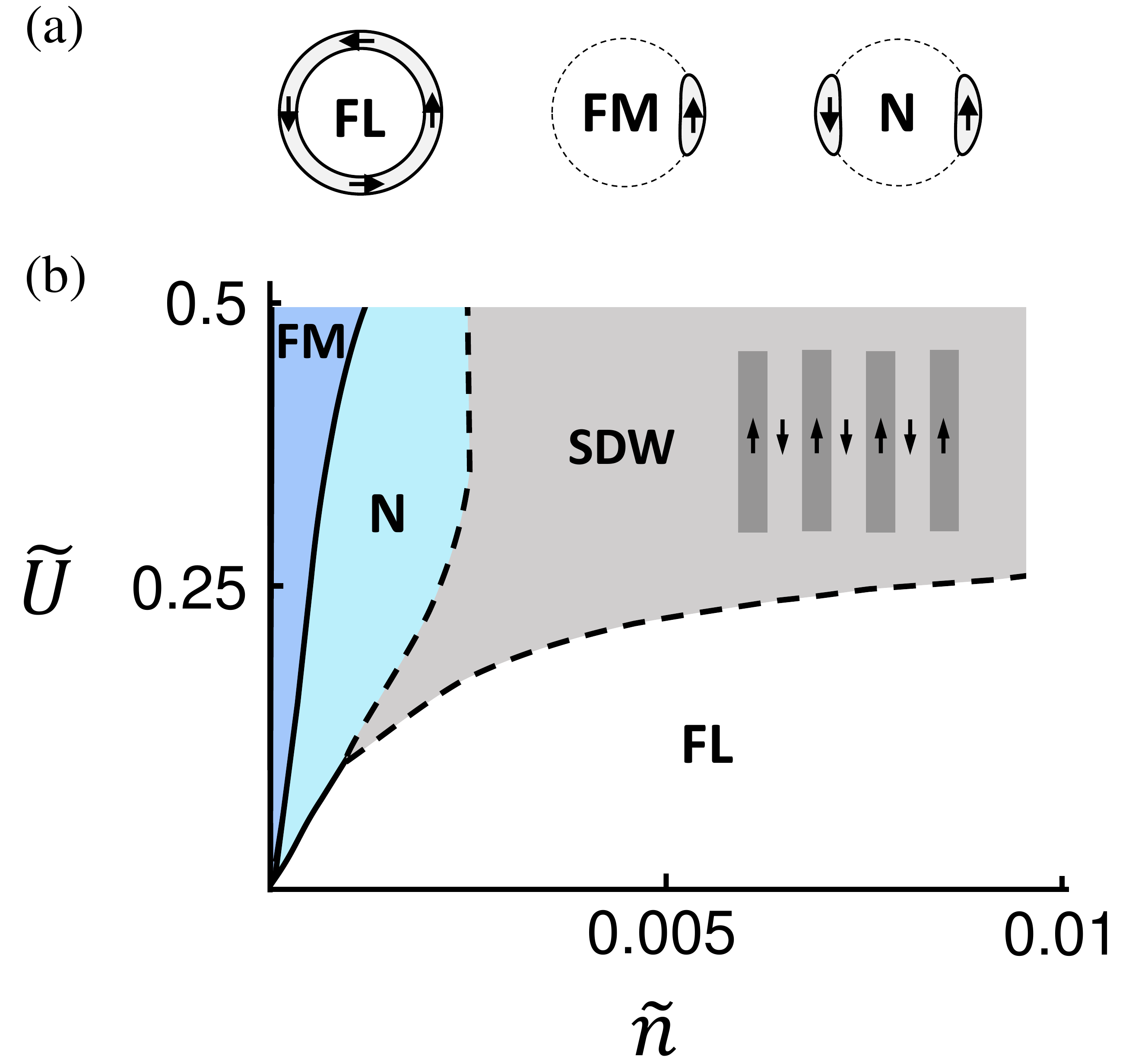}
\caption{(a) The Fermi surface in the Fermi liquid (FL), ferromagnetic (FM), and nematic (N) phases. (b) The phase diagram of the dilute Rashba gas with short range interactions, as a function of the dimensionless density $\tilde n = n/k_0^2$ and dimensionless bare interaction strength $\tilde U = (k_0^2  /2\e_0)\,U$.       }\label{fig:phase_diagram}
\end{figure}

Here, we analyze the fate of an interacting two-dimensional system of fermions with strong Rashba-type spin orbit coupling in the low-density limit. In this limit, the two-particle effective low-energy interaction is strongly renormalized, and obtains a universal form.~\cite{Yang2006,Gopalakrishnan2011}
We analyze the phase diagram, and find a competition between several symmetry-broken liquid states, including a spin-density wave, nematic, and an in-plane ferromagnetic nematic phase (Fig.~\ref{fig:phase_diagram}); in the case of short-range interactions, the ground state in the extreme low-density limit is the ferromagnetic nematic, whereas with interactions that decay as $1/r^3$, where $r$ is the inter-particle distance (which is the case, e.g., for Coulomb interactions screened by a nearby metallic gate, or for dipolar particles with dipole moments pointing perpendicular to the plane), the ground state is a non-polarized nematic.

Finally, we argue that the ferromagnetic and nematic phases are expected to be non-Fermi liquids, due to the strong scattering of quasi-particles near the Fermi surface off the Goldstone modes of the ordered state.~\cite{Oganesyan2001,Metlitski2010,Mross2010}
In the ferromagnetic phase the strong coupling to the magnetic Goldstone modes is generated by spin-orbit coupling~\cite{Xu2010,Bahri2014}.
Rashba spin-orbit coupling thus offers a natural route to realizing a non-Fermi liquid phase.
This phase has been studied extensively in the literature\cite{Oganesyan2001,Metzner2003,Metlitski2010,Markus2010,Mross2010,Xu2010,Mahajan2013,Fitzpatrick2013,Dalidovich2013,Fitzpatrick2014,Bahri2014}; although its nature is still not completely understood, it is believed to be characterized by anomalous power law temperature dependence of physical quantities, such as the specific heat and the resistivity.

Our results are particularly relevant for cold atom experiments. We present an alternative method to study the strongly interacting regime of cold Fermi gases without tuning too close to the Feshbach resonance.~\cite{Ketterle2013} In a spin-orbit coupled gas, the interactions are effectively enhanced due to the large density of states in the low-density limit. This is crucially different from tuning to the Feshbach resonance from the repulsive side, where the decay time to the bound state becomes very short.~\cite{Pekker2011}
Formation of molecules limits the range of accessible interaction strength and has prevented the observation of the ferromagnetic instability thus far.~\cite{Ketterle2013}

This paper is organized as follows. Sec.~\ref{sec:model} describes the model Hamiltonian. In Sec.~\ref{sec:renormalization} we calculate the exact two-particle T-matrix for the case of short ranged interactions. The T-matrix is then used to approximate the effective interactions in the low-density limit. In Sec.~\ref{sec:MF} we numerically compute the phase diagram presented in Fig.~\ref{fig:phase_diagram}.b. Sec.~\ref{sec:dipolar} analyzes the case of dipolar interactions. We then turn to discuss the validity of our results for systems that do not possess perfect rotational symmetry in Sec.~\ref{sec:anisotropy}. Finally, we analyze the effects of the collective mode fluctuations including the stability of the ordered phases to quantum fluctuations and their effect on the lifetime of quasi-particles near the Fermi surface in Sec.~\ref{sec:collective}.

\section{ Model Hamiltonian  } \label{sec:model} We consider a system of fermions in two dimensions with a Rashba-type spin-orbit coupling. The single-particle Hamiltonian is
\be
\mc H_0 =  \sum_{\bs k}\; c_{\bs k}^\dag \hat H(\bs k)\, c^{\vphantom{\dagger}}_{\bs k} \label{H_0}
\ee
 where $\hat H(\bs k) =  {k^2\over 2m }-\mu+\e_0  -  \a ({\bs k}\times \bs \s)\cdot \hat{\bs z}$,
 $\a$ is the strength of the spin-orbit coupling, $\mu$ is the chemical potential, and $\e_0 = m \a^2 /2$ is the spin-orbit energy scale. $c^\dag = (c_{\ua} ^\dag,c_{\da}^\dag)$  is a two component spinor and $\bs \s $ is the vector of Pauli matrices in the same basis. Since we are interested in the low density limit, $\mu \ll \e_0$, we will consider only the low energy band, whose dispersion is
\be
\e_{\bs k} = {\e_0} \left(k- k_0 \right)^2/ k_0^2\,,\label{dispersion}
\ee
where $k_0 = m \a$ is the radius of the circular dispersion minimum. The annihilation operator for a particle in this band is
$
\psi_{\bs k} =\left( c_\ua + i\,e^{i \phi_{\bs k}} c_{\da}\right){/\sqrt 2},
$
where $ \phi_{\bs k} \equiv \arctan \left({k_y / k_x}\right)$ is the angle of the vector $\bs k$ (which is perpendicular to the spin direction). For $\mu<\e_0$ the Fermi sea has the topology of an annulus, with two concentric Fermi surfaces at $k = k_0 \pm k_F$, where $k_F = \sqrt{2 m \mu}$. The single particle density of states is
\[\rho(\mu)  ={\rho_0}\, {k_0 \over k_F}\,,\]
where $\rho_0 \equiv m/\pi$.

The fermions interact via a two-particle repulsion. We will focus on two physically relevant cases: short range (contact) interactions, which are natural in the context of cold atomic gases, and dipolar interactions that decay as $1/r^3$, occurring in two-dimensional electron gases with a nearby screening metallic gate. For simplicity, we consider contact interactions first. The interaction Hamiltonian projected to the lower band is written as
\begin{align}
\mc H_I = {1\over 4 \Omega }\sum_{\bs k \bs k' \bs Q} \, \G_0 (\bs k ,\bs k';\bs Q) \,\,\psi_{\bs k+\bs Q}^\dag \psi_{\bs k'-\bs Q }^\dag \psi^{\vphantom{\dag}}_{\bs k' } \psi^{\vphantom{\dag}}_{\bs k }\,,\label{contact_int}
\end{align}
where $\G_0(\bs k,\bs k';\bs Q) =U\,e^{i\phi_{\bs{k},\bs {k+Q}}}$ and $U$ is the strength of the interaction, $\Omega$ is the volume, $\phi_{\bs{a},\bs {b}} \equiv \phi_{\bs a} - \phi_{\bs b}$, and the factor of $e^{i\phi_{\bs k ,\bs k+\bs Q}}$ arises from the projection to the lower spin-orbit band. More extended dipolar interactions will be considered later.

\section{ Renormalization of the two-particle vertex  } \label{sec:renormalization} We now turn to discuss the renormalization of the two-particle interactions in the case of a circular dispersion minimum. The derivation of the renormalized interaction proceeds along similar lines to that of Ref.~\onlinecite{Yang2006}.

\begin{figure}
\centering
\includegraphics[width=8.5cm,height=4cm]{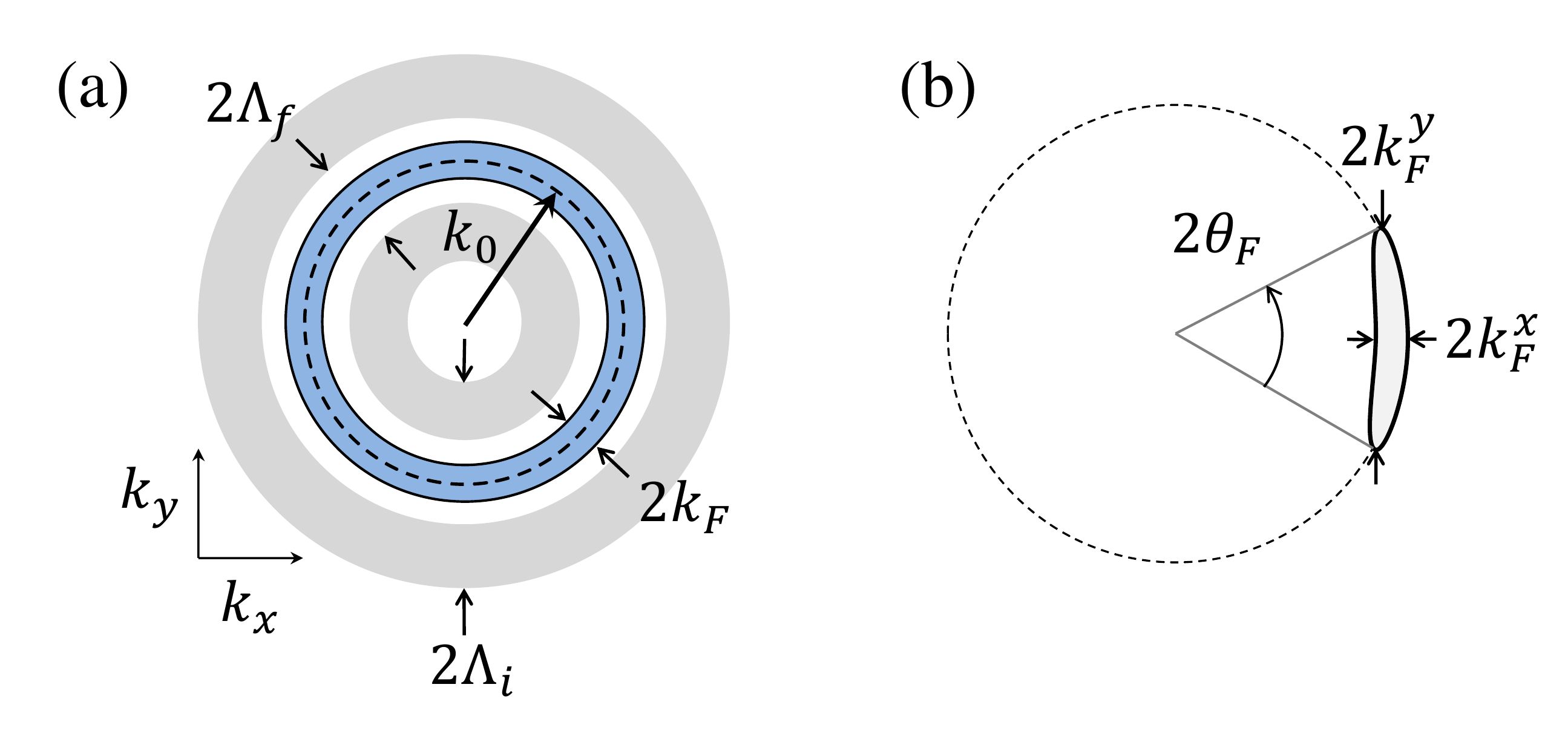}
\caption{ (a) The Fermi sea in the Fermi liquid state defined by the radii $k = k_0 \pm k_F$ and the two high energy shells at $k_0-\L_i<k<k_0-\L_f$ and $k_0+\L_f<k<k_0+\L_i$. (b) Fermi sea in the FM phase. The Fermi surface is highly anisotropic characterized by the Fermi wavelength in the radial direction $k_F ^x$ and in the azimuthal direction $k_F^y \approx k_0 \theta_F$.}\label{fig:spectrum}
\end{figure}

\begin{figure}
\centering
\includegraphics[width=7.5cm,height=6.5cm]{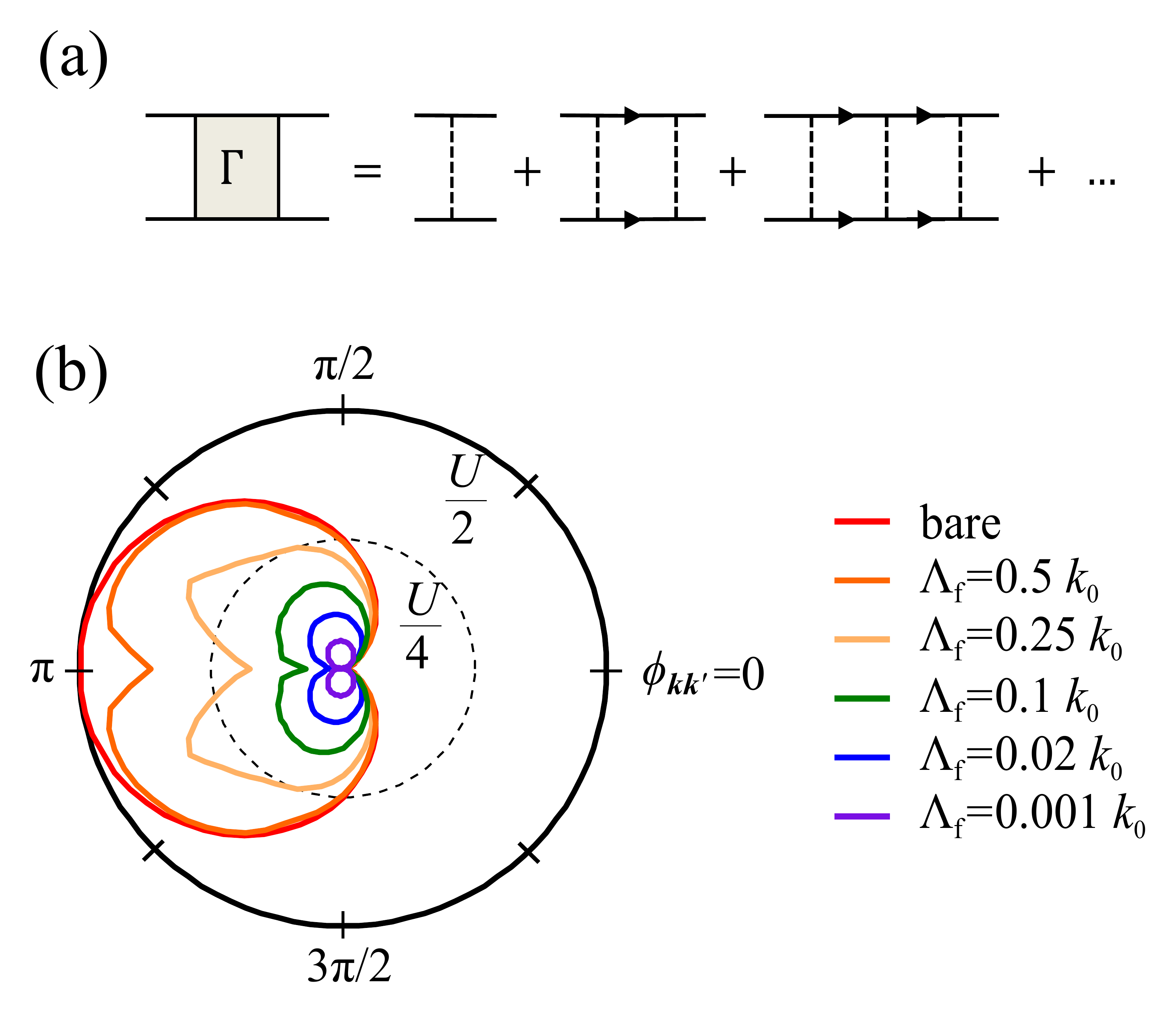}
\caption{(a) A diagrammatic representation of the ladder series for the T-matrix. (b) The angular dependance of the effective interaction (\ref{eff_int_con}) for different values of the lower momentum cutoff $\L_f$.   }\label{fig:renormalized_interactions}
\end{figure}

 We are  interested in the corrections to the bare interaction vertex (\ref{contact_int}) generated by integrating out high energy virtual states, which lie in the two momentum shells $k_0-\L_i<q<k_0 - \L_f$ and $k_0 + \L_f<q<k_0 + \L_i $ (see Fig.~\ref{fig:spectrum}.a). Here $\L_i$ is the high momentum cutoff (which is initially taken to be of order $k_0$) and $\L_f$ is the low momentum cutoff (which is of order $k_F$). This procedure resembles the momentum shell renormalization group approach for fermions,\cite{Shankar1994} except for the fact that here we are integrating out empty states at energies greater than the Fermi energy $\mu$. In this case only the Bardeen-Cooper-Schrieffer diagram\cite{Shankar1994} contributes (see Fig.~\ref{fig:renormalized_interactions}.a). Summing all ladder diagrams we obtain the two particle T-matrix
\be
\G (\w,\bs k,\bs k' ;\bs Q) =   {\G_0(\bs k,\bs k';\bs Q) \over 1 +\mc B(\w,\bs {P})U },\label{eff_int_con_full}
\ee
where $\bs P = \bs k + \bs k'$, { $\omega$ is the sum of the frequencies of the incoming particles, $\bs Q$  is the momentum transfer in the scattering}, and 
$
\mc B(\w,\bs P)=\int_{d \L} {d^2 q\over (2\pi)^2} {1-e^{i\phi_{\bs q \bs{P-q}} } \over -i \w +\xi_{\bs q}+\xi_{\bs P-\bs q}}\,.~\label{B}
$
Here, $d\L$ denotes integration over the regions where both ${\bs q}$ and ${\bs P-\bs q}$ belong to the shells that are integrated out, and $\xi_{\bs k} \equiv \e_{\bs k} - \mu$.

In the dilute limit, $k,k' \simeq k_0$ such that $P \simeq 2 k_0 |\cos {\phi_{\bs k ,\bs k'}\over 2}|$. The renormalized forward scattering interaction {($\bs Q = 0$ and $\bs Q = \bs k' - \bs k, \omega=0$)} assumes the form
\be
V(\phi_{\bs k ,\bs k'}) = { {2}\sin^2 {\phi_{\bs k ,\bs k'} \over 2} \over {1\over U} + \mc B(0,2 k_0 |\cos {\phi_{\bs k ,\bs k'} \over 2}|) } \label{eff_int_con}
\ee
The angular dependance of the interaction (\ref{eff_int_con}) for different values of $\L_f$ is presented in Fig.~\ref{fig:renormalized_interactions}.b. We identify two important features.
First, for forward scattering ($\phi_{\bs k ,\bs k'} \sim 0$) the interaction vanishes as $V\sim U\phi_{\bs k,\bs k'}^2/2$ for all values of $\L_f$. This is because of the Pauli exclusion between spin states with equal orientation.
Second, a strong renormalization occurs 
when the two incoming momenta have opposite directions ($\phi_{\bs k ,\bs k'} \sim \pi$), where the bare interaction is maximal. The strong renormalization results from the large phase space for scattering into the high energy shells when $\bs P \simeq 0$.

One can write analytic expressions for the renormailzed interactions near the points $\phi=0,\pi$\cite{Yang2006}.
As mentioned above, in the case of $\phi_{\bs k, \bs k'} \ll 1$,
\be
V(\phi_{\bs k, \bs k'} \simeq 0 )  \approx {U \over 2 }\left(\phi_{\bs k, \bs k'}\right)^2\,. \label{F}
\ee
On the other hand, for ($\phi_{\bs k, \bs k'}\simeq \pi$) and $\L_f \ll k_0$, the effective interaction assumes a universal form
\be
V(\phi_{\bs k, \bs k'} \simeq \pi)\approx {\L_f \over 4 \rho_0 k_0 K\left(-{ P^2 \over 2 \L_f ^2} \right) }, \label{C}
\ee
where $K(x) = \int_0^{\pi/2} dx (1-x\cos^2 x)^{-1/2}$ is the complete elliptic integral of the first kind, which decays as $K(-x) \approx {\log x \over \sqrt{8 x}}$ for $x\rightarrow \infty$.

\section{ Mean-field phase-diagram  }\label{sec:MF}
To obtain the zero temperature phase diagram (Fig.~\ref{fig:phase_diagram}.b) we use a mean-field approximation with the renormalized interactions (\ref{eff_int_con_full}). First, we compare the energy of two uniformly ordered (translationally invariant) trial states: the ferromagnet (FM) and nematic (N) phase. We then check the stability of these phases towards a spin-density wave state (SDW).

Let us start from the uniform phases. The Fermi surfaces of the FM and N phases presented in Fig.~\ref{fig:phase_diagram}.a are naturally favored by the angular dependance of the renormalized interaction (\ref{eff_int_con}). This is because the interaction is minimal at $\phi_{\bs k \bs k'} = 0$ and $\phi_{\bs k \bs k'}=\pi$ and therefore quasi-particles pairs have the lowest interaction energy when their momenta are collinear.
The FM state is obtained by confining the particles to a finite segment of the ring centered around a specific direction in momentum space, for example $ \hat {\bs k} = \bs{\hat x}$ (as in Fig.~\ref{fig:spectrum}.b). The spin, which is locked perpendicular to the momentum direction, has a non-zero average value. As a result the FM phase breaks time-reversal and rotational symmetry. The N state is obtained similarly by confining the fermions to two such Fermi surfaces residing on two opposite sides of the degeneracy ring. In this case the spin-density is zero on average, and therefore this state breaks only rotational symmetry.

To compare the ground state energy of the FM and N states we expand the interaction (\ref{eff_int_con}) in Fourier components \be
V(\phi_{\bs k \bs k'}) = \sum_{l} V_l e^{i l\left(\phi_{\bs k} - \phi_{\bs k'} \right)}\,.~\label{V_Fourier}
\ee
The trial wave functions are generated by the mean-field Hamiltonian
\begin{align}
\mc H_{MF}^N=\mc H_0 - \sum_{\bs k l} h_l \cos (l \phi_{\bs k}) \,n_{\bs k} \,, \label{MF_nematic}
\end{align}
where we will restrict ourselves only to $l=1,2$ solutions (which are ferromagnetic and nematic, respectively). Minimizing the expectation value of the full Hamiltonian with respect to $\mu$, $h_1$ and $h_2$ yields the equations (see Appendix \ref{app:var})
\begin{align}
&n = {1\over \Omega}\sum_{\bs k}\langle n_{\bs k}\rangle, \label{VE1} \\
&h_1 = -{2V_1\over \Omega }\sum_{\bs k} \cos{ \phi_{\bs k}}\,\langle n_{\bs k}\rangle, \label{VE2}\\
&h_2 = -{2V_2\over \Omega }\sum_{\bs k} \cos{2 \phi_{\bs k}}\,\langle n_{\bs k}\rangle \label{VE3}\,.
\end{align}
The FM state is characterized by $h_1 \ne 0$, while the N phase corresponds to $h_2 \ne 0$ and $h_1 = 0$.
At sufficiently low density, the ground state is always the FM state; upon increasing the density, there is a first-order transition to a N state, followed by another first-order transition to a rotationally invariant FL state. The fact that the transitions are of first order is associated with the presence of a nearby van Hove singularity in the density of states.~\cite{Fischer2013}

We now turn to discuss the stability of the uniformly ordered states towards textured phases (either spin or charge density waves).
First we note that in the low-density limit the Fermi surface contains nearly nested segments which are separated by $q=2 k_F$, where $2k_F$ is the Fermi wavelength along the radial direction.
As a result, the charge and spin susceptibility  $\chi_{\rho,\s}(\bs q)$ are sharply peaked at $q=2k_F$\cite{Yang2006} (see Appendix \ref{app:spin susceptability}).
For sufficiently short-range interactions, the FM phase is always stable to SDW and CDW formation in the low density limit. This is because the system is nearly spin polarized, and the interaction between fermions on the Fermi surface is small.

The FL and N phases become unstable to SDW formation when the Stoner criterion $V_1 \chi_\perp (q) = 1$ is satisfied, where $\chi_\perp(q)$ is the in-plane spin-susceptibility transverse to $\bs q$ (see Appendix~\ref{app:spin susceptability}). 
The transition lines to the SDW phase are shown as dashed lines in Fig.~\ref{fig:phase_diagram}.

In the low density limit, the angular size of the Fermi surfaces in the FM and N phases becomes small. We can then utilize the asymptotic analytic expressions for the effective interaction near $\phi_{\bs k, \bs k'} = 0, \pi$ [Eqs. (\ref{F},\ref{C}) and Ref.~\onlinecite{Yang2006}] to estimate the ground state energy. 
The shape of the Fermi surfaces is highly anisotropic, $k_F ^x \ll k_F ^y$, where $k_F^x$ ($k_F ^y$) is the Fermi wavelength along the radial (azimuthal) direction (see Fig.~\ref{fig:spectrum}.b).

In the N phase, the Fermi surface consists of two such anisotropic patches.
In this case, the inter- and intra-patch interactions are given by (\ref{F}) and (\ref{C}), respectively. (The lower cutoff for the renormalization procedure of the interaction is taken to be $\L_f = 2k_F^x$.)
The total momentum $P = |\bs k + \bs k'|$, which enters the inter-patch interaction (\ref{C}), is much greater than $k_F ^x$, over most of the Fermi surface. We can therefore use the approximate form of (\ref{C}) for $P\gg \L_f $:
\be
V(\phi_{\bs k\bs k'}\simeq \pi) \approx{|\phi_{\bs k \bs k'}-\pi| \over 4 \rho_0 \log {k_0 / k_F ^x} }\,. \label{C4}
\ee
On the other hand, the intra-patch interaction (\ref{F}) decays quadratically at small angles.

The total energy per particle in the N phase scales as
\be\e_N \propto\left({ \e_0 \over \rho_0^2 k_0 ^4}\right)^{1/3}{ n^{4/3}\over \left(\log {k_0^2 \over n}\right)^{2/3}}\,. \label{EN}\ee
while in the FM phase the energy per particle is
\[\e_{FM} \propto {\sqrt{U \e_0 }\over k_0^2}n^{3/2}.~\]
We conclude that for short-ranged interactions in the zero density limit, the ground state is FM, in agreement with Ref.~\onlinecite{Berg2012}.

\section{ Dipolar interactions  } \label{sec:dipolar}
We now turn to discuss the case of dipolar interactions, which decay as $1/r^3$ at large distances. In Fourier space, the interaction is given by $U_{d}(\bs q) \approx v_1-v_2 q$ for small $q$. The corresponding bare interaction vertex assumes the form
\begin{align}
\mc H_I^d ={1\over \Omega} \sum_{\bs{p,p',P}} \G_0^{d}(\bs p ,\bs{p'};\bs P) \psi_{\bs{p'}}^\dag \psi_{\bs{P-p'} }^\dag \psi^{\vphantom{\dag}}_{\bs{P-p}} \psi^{\vphantom{\dag}}_{\bs p }\,,\label{dd_int}
\end{align}
where the vertex function is given by
\begin{align}
\G_{0}^{d}(&\bs p,\bs {p'};\bs P) = U_{d}(|\bs{p-p'}|)\times \label{G_dd} \\&\left[ 1+e^{i \phi_{\bs p,\bs{p}'}} +e^{i \phi_{\bs {P-p},\bs{P-p}'}}+e^{i \phi_{\bs {p},\bs{p}'}+i\phi_{\bs {P-p},\bs{P-p}'}} \right].~\nn
\end{align}
The one-loop correction to the effective interaction then takes the form
\begin{align}
\mc B_{d}(\w,\bs P)=\int_{d \L}& {d^2 q\over (2\pi)^2}\bigg[ {{\G_0^{d}(\bs p,\bs {q};\bs P)\G_0^{d} (\bs q,\bs {p}';\bs P)} \over -i \w +\xi_{\bs q}+\xi_{\bs P-\bs q}}\label{B_dd}\\&-{\G_0^{d}(\bs p,\bs {q};\bs P)\G_0^{d} (\bs{P-q},\bs {p}';\bs P)\over -i \w +\xi_{\bs q}+\xi_{\bs P-\bs q}}\bigg]\,.~\nn
\end{align}

The interaction vertex (\ref{G_dd}), which now depends non-trivially on the momentum transfer $\bs Q = \bs p - \bs p'$, becomes particularly simple in the limits of interest: (i) the Cooper channel ($ P \ll k_0$) and (ii) forward scattering ($P \simeq 2k_0$).
For Cooper channel scattering, case (i), the bare vertex (\ref{G_dd}) becomes a function of a single angle
\begin{align}
\G_{0}^{d}(\phi_{\bs{pp'}}) = U_{d}(\phi_{\bs{pp'}})\left[ 1+e^{i \phi_{\bs p\bs{p'}}}  \right]^2\,. \label{G_0d}
\end{align}
We expand (\ref{G_0d}) in Fourier components $\G_0 ^d(\phi_{\bs p \bs p'}) = \sum_m \G_{0,m} ^d  e^{i\,m\,\phi_{\bs p \bs p'}} $ and insert it into (\ref{B_dd}) to obtain\cite{Yang2006} 
\be
\mc B_{d}(P\simeq0) = -4\rho_0{k_0 \over \L_f} K\left(-{ P^2 \over 2 \L_f ^2} \right)\sum_{m=1}^\infty [1-(-1)^m] (\G_{0,m} ^d ) ^2
\ee
in the $\L_f \rightarrow 0$ limit. Note that $m$ denotes the total angular momentum (orbital plus spin), and that only the odd ones contribute.
The different angular momentum channels decouple in the ladder series (Fig.~\ref{fig:renormalized_interactions}.a) due to conservation of angular momentum, just as in equation (\ref{eff_int_con_full}). In the low density limit they all assume a universal form
\be
\G _{2m+1}^d(\phi_{\bs k\bs k'}\simeq \pi) \approx C_d(\phi_{\bs k\bs k'})\equiv{ |\phi_{\bs k \bs k'} -\pi| \over 4 \rho_0 \log \left( {k_0 \over \L_f} \right) }\,. \label{C2}
\ee
where $\bs k = \bs p$, $\bs k' = \bs P-\bs p$. The total angle dependant interaction then assumes the form
\begin{align}
V_d &(\phi_{\bs k\bs k'}\simeq\pi) =C_d(\phi_{\bs k\bs k'})\times\label{C3}\\
&\lim_{N \rightarrow \infty}{1\over N}\sum_{m=0 }^N\, e^{i \,(2m+1)\,\phi_{\bs p \bs p'}} = C_d(\phi_{\bs k\bs k'})\,\d_{\phi_{\bs p \bs p'} ,0}\nn
\end{align}
Overall for Cooper channel scattering we obtain the same result as for short-ranged interactions (\ref{C}).

In the case of forward scattering, $P\simeq 2k_0$, the integrand of (\ref{B_dd}) diverges at $\bs q = \bs p$ and $\bs q = \bs P-\bs p$. The integral is dominated by the vicinity of these two points, whose most divergent contribution as $\Lambda_f \rightarrow 0$ is
\be
\mc B_d(\phi_{\bs k \bs k'})\approx {\left[\tilde U _d (\phi_{\bs k \bs k'})\right]^2 \over |\sin \phi_{\bs k \bs k'}|}\rho_0 \log \left(k_0 \over \L_f \right)
\ee
where $\tilde U_d (\phi_{\bs k \bs k'}) \equiv 2U_d(0)-(1+\cos \phi_{\bs k \bs k'} )U_d (2k_0 |\sin {\phi_{\bs k \bs k'} \over 2}|)$ which decays linearly near $\phi_{\bs k \bs k'}=0$
$\tilde U_d(\phi_{\bs k \bs k'})\approx 2 v_2 k_0|\phi_{\bs k,\bs k'}| $.
Summing up the ladder series we have
\[V_{d}(\phi_{\bs k \bs k'} \simeq 0) = {\tilde U_d (\phi_{\bs k \bs k'}) \over 1+ \mc B_d (\phi_{\bs k \bs k'})\tilde U_d (\phi_{\bs k \bs k'})}.\]
Therefore,
just as in the case of short-ranged interactions, we recover the bare interactions for small angle scattering ($\phi_{\bs k \bs k'}\simeq 0$):
\be
V_{d}(\phi_{\bs k , \bs {k'}}\simeq0 )\approx 2 v_2 k_0|\phi_{\bs k,\bs k'}|\,.~\label{F2}
\ee

Using the asymptotic form of the effective interaction, Eqs. (\ref{C3}) and (\ref{F2}), we can estimate the ground state in the zero density limit, just as we have for short ranged interactions at the end of section \ref{sec:MF}.
The crucial difference is that now the forward scattering term (\ref{F2}) decays linearly to zero near $\phi_{\bs k \bs k'}\simeq 0$  and not quadratically as it did for short-ranged interactions (\ref{F}). As a result, the energy of the FM phase scales as $\e_{FM} \propto n^{4/3}$, whereas the scaling of the energy of the N phase is unmodified compared to contact interactions [Eq. (\ref{EN})].
Therefore, we conclude that for dipolar interactions, the ground state in the zero density limit is the nematic state, due to the logarithm in Eq.~(\ref{EN}). This connects to the results of Ref.~\onlinecite{Berg2012}, which predicted that for interactions that decay like $1/r^a$ the value $a=3$ is critical, separating between the anisotropic Wigner crystal (AWC) and the FM. The N phase can be viewed as a melted version of the AWC phase.

\section{ Effects of breaking of the rotational symmetry  } \label{sec:anisotropy}
Most physical realizations of the dispersion (\ref{dispersion}) will include additional terms which break the rotational symmetry.
In solid state systems such terms arise from the underlying lattice, while in cold atom systems they are due to the Raman
lasers.~\cite{Campbell2011}
To study the effects of these terms on our results, we add the symmetry breaking term
\be\mathcal{H}_\epsilon = -\a( \epsilon-1) k_y \s^y .~\label{Heta}\ee
to the Hamiltonian (\ref{H_0}).
Here $\epsilon\leq 1$ is a parameter that describes the degree of a two-fold anisotropy ($\epsilon=1$ corresponds to the isotropic case). In this case, the low-energy helical quasi-particles are given by $\psi_{\bs k} =
\left(c_{\bs k \ua}+i\,e^{i \tilde \phi_{\bs k}} c_{\bs k \da} \right)$, with $\tan \tilde \phi_{\bs k} = \epsilon k_y / k_x$.
We calculate the effect of the symmetry breaking term on the solution of the self-consistency equations
(\ref{VE1}-\ref{VE3}) for the case of a FM transition.

\begin{figure}
\centering
\includegraphics[width=6.cm,height=4.8cm]{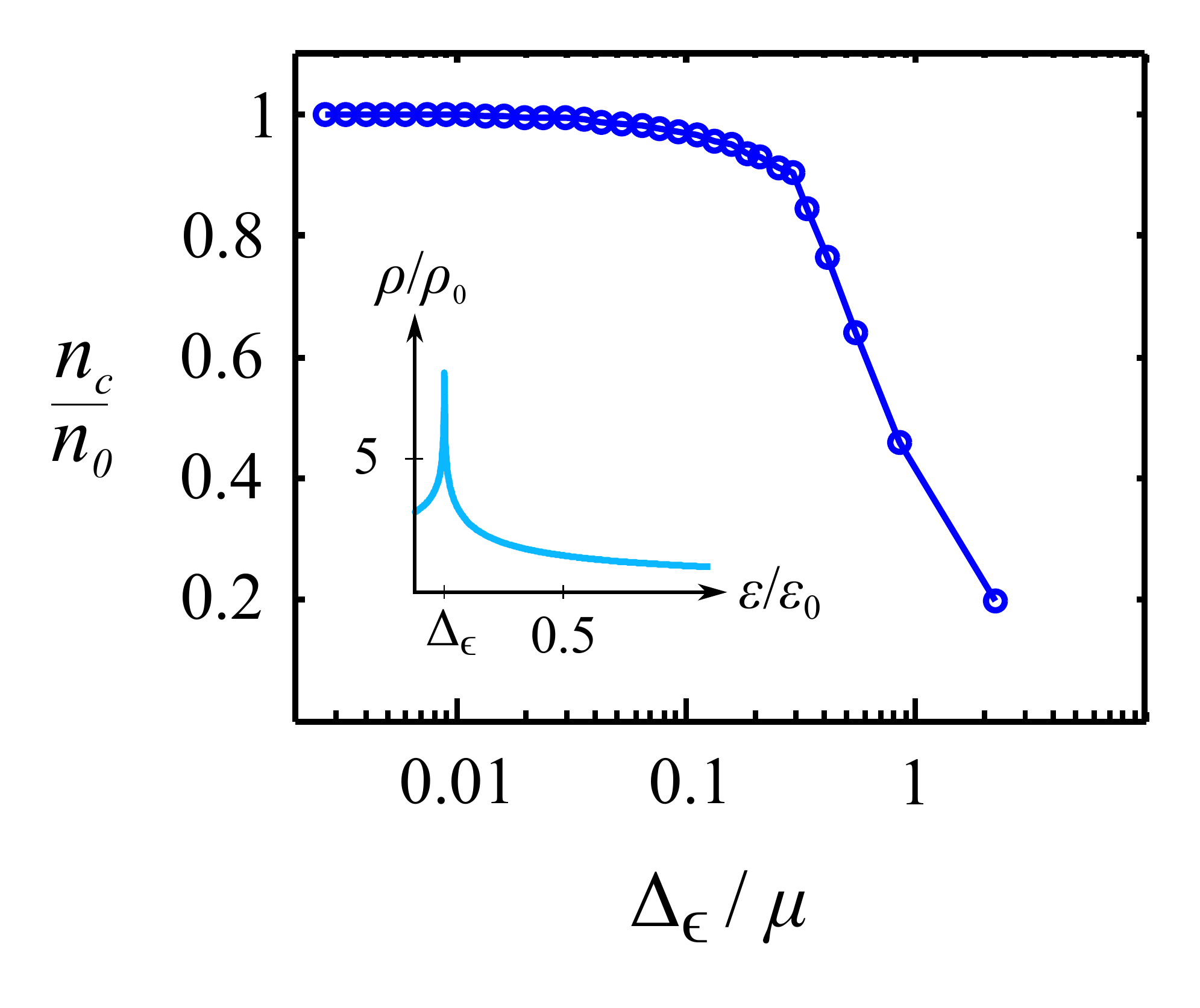}
\caption{The critical density, $n_c$, of the ferromagnetic transition for $\tilde U = 1.5$  in the presence of anisotropy in the single-particle dispersion (Eq.~\ref{Heta}) normalized by the critical density without
the anisotropy term ($\epsilon = 1$), $n_0$, vs. the anisotropy energy scale $\D_\epsilon\equiv\e_0(1-\epsilon^2) $ normalized by the chemical potential
at the transition. 
The inset shows the
density of states as a function of energy for the anisotropic Rashba dispersion.  }\label{fig:anisotropy}
\end{figure}

 Before discussing the results, we note that the symmetry breaking term (\ref{Heta}) does not modify the
 renormalization of interactions presented in section \ref{sec:renormalization} as long as the Fermi energy is much greater than the
 energy scale associated with the anisotropy, $\D_\epsilon = \e_0(1-\epsilon^2)$. In this limit the Fermi sea of the non-interacting gas still
 has the form of an annulus with a density of states which increases with decreasing density (see inset of Fig.~\ref{fig:anisotropy}).
In the opposite limit, $\mu\ll \D_{\epsilon}$, the Fermi surface is broken into two elliptic surfaces (similar to the
Fermi surfaces of the N phase in Fig.~\ref{fig:phase_diagram}.a) and the density of states decreases with decreasing density and
chemical potential. In this limit we expect that the renormalization of interactions will be closer to that of a Fermi gas without
spin-orbit coupling.~\cite{Kanamori1963}
In the regime $\mu \gg \D_{\epsilon}$, the renormalized interactions are approximated by
\[V(\phi) \approx{ \sin^2{\tilde \phi_{\bs k \bs k'}\over 2}\over {1\over U} + B(0,P)}\,,\]
where $\bs P = \bs k + \bs k'$ and $\bs k$, $\bs k'$ lie on the elliptic dispersion minimum.
Therefore, we can decouple the interaction in the same way we did in (\ref{MF_nematic}) and solve using the same self-consistency
equations (\ref{VE1}-\ref{VE3}) with $\tilde \phi_{\bs k}$ instead of $\phi_{\bs k}$.

Fig.~\ref{fig:anisotropy} presents the critical density for the transition into the FM phase normalized by the critical density at $\epsilon=1$ vs.
the anisotropy energy $\D_\epsilon$ divided by the chemical potential at the transition. We find that the symmetry breaking term $\mc
H_\epsilon$ has a negligible effect when the transition occurs at $\mu \gg \D_\epsilon$. However, when $\D_\epsilon$ approaches $\mu$ at the
transition, the critical density drops rapidly, and the FM order is obstructed by the anisotropy.

\section{ Collective excitations and non-Fermi liquid behavior } \label{sec:collective}
We now turn to discuss the effects of fluctuations of the order parameter in the FM and N phases. These phases break the (continuous) rotational symmetry of the system. The resulting gapless Goldstone mode associated with this symmetry breaking is strongly coupled to the quasi-particles at the Fermi
energy.~\cite{Oganesyan2001,Metlitski2010,Xu2010}
This coupling gives rise to two important effects: first, the Goldstone modes become Landau damped by the particle-hole
excitations near the Fermi surface. Second, the Landau quasi-particles are strongly scattered by the Goldstone mode,
leading to the break down of the Fermi liquid behavior.~\cite{Oganesyan2001,Metlitski2010,Mross2010,Xu2010,Dalidovich2013,Vishwanath2014}

Below, we use Hertz-Millis\cite{Herz1975,Millis1993} type arguments to demonstrate that such a strongly coupled state indeed arises in the N and FM phases in our setup.
Hertz-Millis theory is known to ultimately fail in $d=2$;~\cite{Metlitski2010,Mross2010,Thier2011} nevertheless, following Ref.~\onlinecite{Oganesyan2001}, we argue its application shows that a Fermi liquid ground state is inconsistent.

We consider, for example, the FM phase with short-ranged interactions. We employ the Hubbard-Stratonovich transformation
to decouple the imaginary time action using the magnetization field $M_q = M_q^y -i M_q^x$:
\begin{align}
\mc S =& \sum_{ k } \psi_{ k }^\dag\left( -i\w +\xi_{ \bs k} \right)\psi_{ k}\nn \\&-  V_1\sum_{\bs k q}\left(e^{i \phi_{ \bs k +
{\bs q \over 2}}} M_{ q}\psi^\dag _{ k + { q \over 2}}\psi _{ k -{ q \over 2}}+c.c.~\right) + V_1\sum_{ q} |M_{ q}|^2\nn
\end{align}
where $q=(\w,\bs q)$ and $k = (\nu,\bs k)$ denote $1+2$ component vectors in frequency and momentum space. We expand the action
around the broken symmetry state $\d M_{ q} = M_q- M  \d_{q,0}$:
\begin{widetext}
\begin{align}
\mc S = \sum_{ k } \psi_{ k }^\dag\left( -i\w +\xi_{ \bs k} ^{FM} \right)\psi_{ k}-  V_1\sum_{\bs k q}\left(e^{i \phi_{ \bs k + {\bs q
\over 2}}} \d M_{ q}\psi^\dag _{ k + { q \over 2}}\psi _{ k -{ q \over 2}}+c.c.~\right) + V_1\sum_{ q} \left(|\d M_q^y- M  \d_{ q
,0}|^2 + |\d M_q^x|^2\right)\label{S}
\end{align}
where $\langle M_q \rangle = M \d_{q,0} $ is taken to be real, and $M = h_1 /V_1$ is given by the solution of the self-consistent equation (\ref{VE2}). The
dispersion of the fermions is given by $\xi_{\bs k}^{FM} = \e_{\bs k} - h_1 \cos \phi_{\bs k} -\mu$.
 The effective Ginzburg-Landau theory for $\d M$ is obtained by integrating out the Fermionic degrees of
 freedom. To second order in $\delta M$ we get
\be
 \mc S_{(2)} = -V_1\sum_{\bs q \w}\left[\d_{ij}-V_1\Pi_{ij}(q)\right]\d M_q ^i \d M_{-q}^j, \label{FGL}
\ee
where the Lindhard function $\Pi_{ij}(q)$ is given by
\begin{align}
\Pi(\bs q,i\omega) \approx\int{ d ^2 k\over (2\pi)^2} {n_F\left(\xi_{\bs k+{\bs q \over
2}}^{FM}\right)-n_F\left(\xi_{\bs k-{\bs q \over 2}}^{FM}\right)\over i \w  - \bs v({\bs k})\cdot \bs q }\,
  { \begin{pmatrix}\sin \phi_{\bs k + {\bs q \over 2}} \sin \phi_{\bs k - {\bs q \over 2}} &  \sin \phi_{\bs k + {\bs q \over 2}} \cos \phi_{\bs k - {\bs q \over 2}} \\  \cos \phi_{\bs k + {\bs q \over 2}} \sin \phi_{\bs k - {\bs q \over 2}} & \cos \phi_{\bs k + {\bs q \over 2}} \cos \phi_{\bs k - {\bs q \over 2}}   \end{pmatrix}} \,.~\label{Pi}
\end{align}
Here $n_F$ is the Fermi function, $ v_i({\bs k}) = 2{\e_0 \over k_0} \left({k \over k_0}-1\right) \hat{ k}_i -{h_1 \over k_0} \, \varepsilon_{ij}\hat{ k}_y{\hat k_j} $ where $\varepsilon_{ij}$ is the antisymmetric tensor, and $\hat {\bs k} \equiv (\cos \phi_{\bs k},\sin \phi_{\bs k})$. To lowest order in $(\omega, \bs q)$ (assuming that $\omega \ll q$) the Lindhard function can be written as
\be
\Pi(\bs q,\w) \approx \begin{pmatrix} {1\over \D_x} - \eta_x(\phi_{\bs q}){|\w|\over q} -\kappa_x q^2 -\tilde \kappa_x(q_x ^2- q_y^2)  & \g q_x q_y \\ \g q_x q_y  & {1\over \D_y} - \eta_y(\phi_{\bs q}){|\w|\over q} -\kappa_y q^2 +\tilde \kappa_y(q_x ^2- q_y^2)\end{pmatrix}. \label{Pi2}
\ee
\end{widetext}
Here, $\eta_{x,y}(\phi_{\bs q})$ are the (direction dependent) Landau damping coefficients, $\kappa_{x,y}$, $\tilde{\kappa}_{x,y}$, and $\gamma$ describe the stiffness of the order parameter to slow spatial modulations, and $\Delta_{x,y}$ determine the gaps of the collective modes. The anisotropy in the parameters of $\Pi(\bs q, \omega)$ is due to the fact that we are working in an ordered phase that breaks rotational invariance. We have determined the parameters by numerically integrating Eq.~(\ref{Pi}) [$\eta_{x,y}$ can also be expressed analytically - see Eq.~(\ref{eta}) below]. We find that $\D_x = V_1$, such that transverse fluctuations of the order parameter are gapless, as required from Goldstone's theorem (the order parameter is assumed to point along the $y$ axis).

From the effective action (\ref{FGL}) we can compute the zero point fluctuations of magnetization field. Deep in the ordered phase, these are dominated by the transverse fluctuations. The deviation of the angle of the order parameter from the $y$ direction is $\d \f=\d M^x /M$, and the mean fluctuations in $\d \f^2$ are given by
\begin{align}
\langle \d\f^2 \rangle &\equiv \int {d\w d^2q\over (2\pi)^3}\langle \d\f_{q}\d\f_{-q} \rangle  \label{f^2}\\&\approx \int {d\w d^2q\over (2\pi)^3}  {1/ h_1^2 \over \eta_x (\phi_q) {|\w | \over q} +\kappa_x q^2 +\tilde \kappa_x (q_x^2 - q_y^2)}\nn
\end{align}
where we have kept only the most singular contribution in the long-wavelength, low-frequency limit, and used $h_1=V_1 M $. $\eta_{x}(\phi_{\bs q})$ can be expressed as
\be
\eta_x(\phi_{\bs q}) \approx \sum_j { k_{0,j} \over (2\pi)^2 v_j^2}  \sin^2(\phi_{\bs k_j}). \label{eta}
\ee
The sum runs over the points $\bs k_j$ on the Fermi surface where the Fermi velocity $\bs v_j$ is perpendicular to $\bs q$. $k_{0,j}$ is the radius of curvature of the Fermi surface at these points.
As a crude approximation, we replace $\eta_x(\phi_{\bs q})$ and the order parameter stiffness in Eq.~(\ref{f^2}) by the average values, $\eta$ and $\kappa$, respectively. (For $\tilde U=0.5$ and $n/k_0^2=10^{-5} - 10^{-3}$ we found that these anisotropies are numerically small.) Eq.~(\ref{f^2}) then assumes the simple form
\be
\langle \d \f ^2 \rangle \approx {1\over 3(2\pi)^2}{Q^3 \over h_1^2 \eta}\log\left(1+{\eta \Omega \over \kappa Q^3} \right),
\ee
where $Q\sim k_F^x$ and $\Omega $ are the momentum and frequency ultraviolet cutoffs, respectively.
Interestingly in our numerically determined values for $k_F ^x$, $h_1$ and $\eta_x$ we find that $\langle \d\f^2 \rangle  \ll1$ for a broad range of interaction strengths and densities. Therefore, we conclude that the FM order is stable against quantum fluctuations in the range of densities $n/k_0^2=10^{-5} - 10^{-3}$, where we have computed the phase diagram Fig.~\ref{fig:phase_diagram}.b.

\begin{figure}
\centering
\includegraphics[width=6.cm,height=11cm]{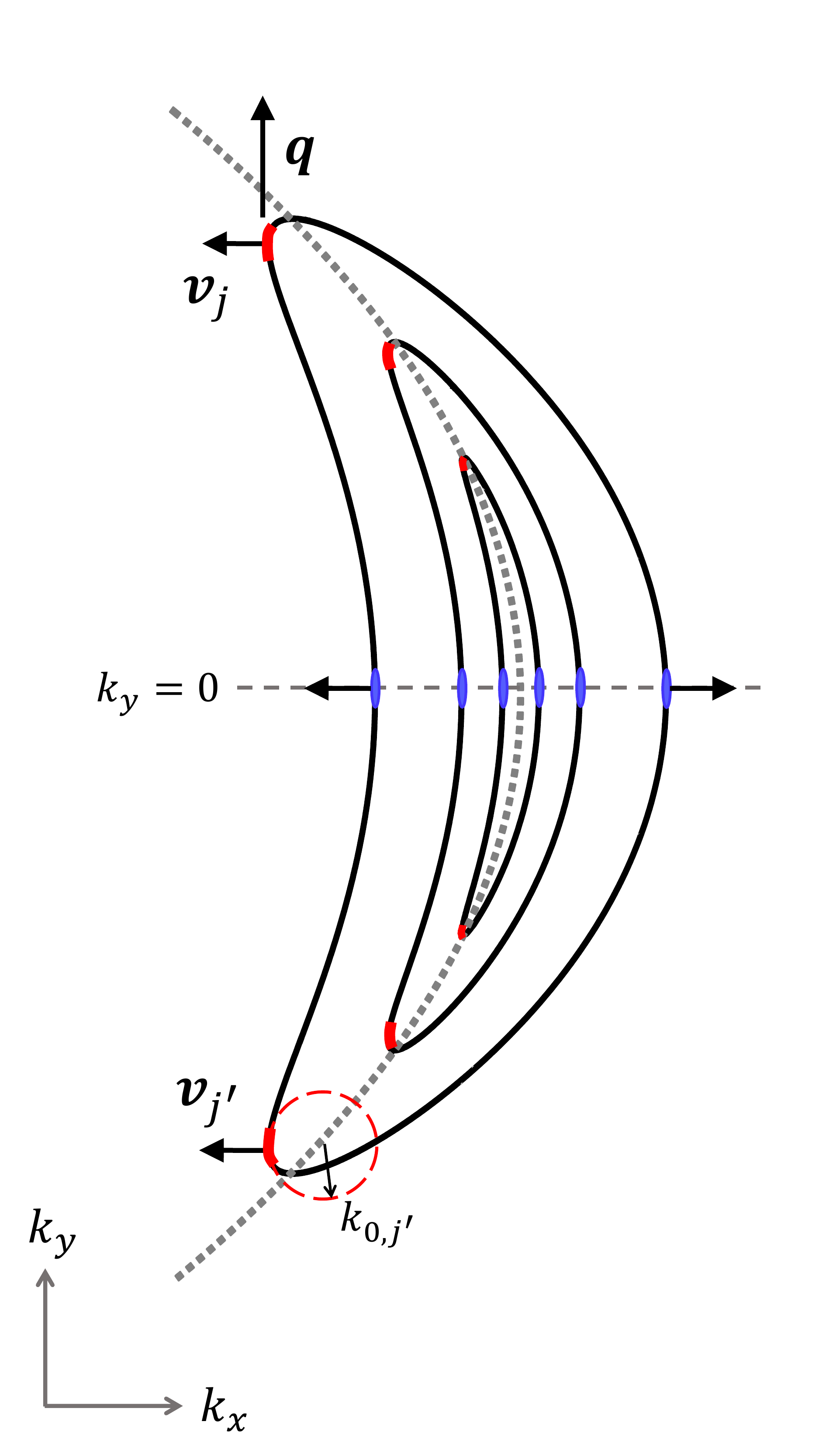}
\caption{ The Fermi surface in the FM phase for three different densities. The red lines denote regions in which $\bs v(\bs k)$ points along the $x$ direction and perpendicular to $\bs q$ which is taken to be along $y$. These points give rise to the Landau damping term in Eq. (\ref{Pi}). The blue lines mark similar points where $\bs v(\bs k)$ is aligned along $x$ but the form factor $\sin\phi_{\bs k}$ in Eq.~(\ref{S}) vanishes. The arrows denote the direction of the velocity $\bs v (\bs k)$, which is  perpendicular to the vector $\bs q$. The small dashed circle denotes the radius of curvature $k_{0,j'}$ of the point $k_{j'}$, which appears in the definition of the Landau damping term Eq. (\ref{eta}). }\label{fig:hotspots}
\end{figure}

It is interesting to note that in the N and FM phases discussed here, the Goldstone modes are over-damped in \emph{any} direction of propagation [i.e., $\eta_x(\phi_{\bs q})$ never vanishes]. This is in contrast to the case of a distorted circular Fermi surface (analyzed in Ref.~\onlinecite{Oganesyan2001}), where the Goldstone modes remain under-damped along a discrete set of angles close to $0$, $\pm \pi/2$.
This is because of the banana-like shape of the Fermi surfaces in our case (see Fig.~\ref{fig:hotspots}). On such a Fermi surface, there are points for which $\bs v (\bs k)  \perp \bs q$ and $\sin^2{\phi_{\bs k}}\ne 0$ for any $\bs q$. E.g., for $\bs q \parallel \hat{\bs y}$, there are points where the Fermi surface is parallel to $\hat{\bs y}$ with $k_y \ne 0$  (marked in red in Fig.~\ref{fig:hotspots}). By Eq.~(\ref{eta}), this implies that the Landau damping term $\eta_x$ is never zero.

Finally, we turn to discuss the fate of the low-energy Fermionic quasi-particles in the FM and N phase. These quasi-particles are coupled to the Goldstone mode through the following term [see  Eq.~(\ref{S})]:
\be\mc L_{\f,\psi} = \lam \,\sin \phi_{ \bs k+{ \bs q \over 2}} \d\f_{ q}\, \psi_{ {k+{q\over 2}}}^\dag \psi_{ {k-{q\over
2}}} +c.c\,,\ee
where the coupling strength is $\lam = h_1$.
Following Ref.~\onlinecite{Oganesyan2001} we consider the coupling term perturbatively, and show that it necessarily leads to the breakdown of Fermi liquid behavior on the entire Fermi surface, except for a discrete set of points.
Using (\ref{FGL}), we obtain the following leading order self-energy correction to the fermion propagator:
\be
\Sigma(\omega,\bs k) = {\lam^2\over (2\pi)^3 h_1^2} \int { \sin ^2 \phi _{\bs k}\,d^2 q\, d\nu \over \left(\eta{|\nu |\over q}+\kappa
q^2\right)\left( i(\nu+\w)+\bs v(\bs k) \cdot \bs q\right)},
\ee
where $\bs k$ lies on the Fermi surface. As before, we have replaced $\eta$ and $\kappa$ by their averages over $\phi_{\bs q}$. Integrating over $q$ and $\w$ yields\cite{Oganesyan2001,Metlitski2010,Sachdev2011}
\be
\Sigma (\omega,\bs k)\approx\left({\lam \over h_1}\right)^2 { \sin^2 \phi_{\bs k}\over 8 \pi ^2 \eta v (\bs k)}\,\mrm{sign}\,\w \left(\eta |\omega|\over
\kappa\right)^{2/3} \label{Sigma}
\ee
For small $\omega$, the self-energy becomes dominant over the bare $i \omega $ term in the Fermionic Green's function. This invalidates the perturbative approach, and signals a breakdown of Fermi liquid behavior. E.g., treating (\ref{Sigma}) naively implies that there is no discontinuity on the Fermionic occupancy $n_{\bs k}$ on the Fermi surface, except at the two points where $\phi_{\bs k}=0$.

From (\ref{Sigma}) we can extract a momentum-dependent energy scale where the leading order self-energy correction becomes comparable to $|\omega|$: $E_0({\bs k}) = \frac{\sin^6(\phi_{\bs k})}{8\pi^2 \eta v^3({\bs k}) \kappa^2}$. At this energy scale, a crossover from Fermi liquid to non-Fermi liquid behavior occurs. Using our numerically obtained values of $\eta$, $v$, and $\kappa$ at the tip of the banana-shaped Fermi surface, we find that this energy scale is always much greater than the Fermi energy. This implies that near the tips, there is no observable Fermi-liquid regime. Near the $\phi_{\bs k}=0$ ``cold spots'', on the other hand, the non-Fermi liquid scale vanishes rapidly as $\sin^6 \phi_{\bs k}$.

In the presence of a weak anisotropy in the dispersion, as in Eq.~(\ref{Heta}), the Goldstone modes are ultimately gapped at low energy. At this energy, another crossover occurs, and at asymptotically low energies Fermi liquid behavior is recovered.

\section{Discussion} \label{sec:discussion}
In conclusion, we have analyzed the fate of a Rashba spin-orbit coupled Fermi gas in the low density limit. The Fermi liquid state is unstable towards a
variety of competing liquid crystalline phases: ferromagnetic-nematic, nematic, and spin-density wave states.
In the case of short-ranged interactions, a cascade of phase transitions occurs as the density decreases. The high density isotropic Fermi liquid undergoes a transition to a spin density wave, followed by a nematic state, and finally the ground state becomes the ferromagnetic-nematic state at asymptotically low densities. In the case of dipolar interactions, the nematic state is the ground state all the way to the zero density limit.

We have also analyzed the stability of the ferromagnetic-nematic phase against terms that break the
rotational symmetry, e.g., due to the underlying crystalline lattice. We found that the ferromagnetic order is stable as long as the chemical potential
at the transition is greater than the energy scale associated with the symmetry breaking term.

Finally, we have discussed the effects of long-wavelength fluctuations of the FM and N order parameter on the low-energy quasi-particles. Scattering off these fluctuations gives rise to the breakdown of Fermi liquid theory, as generally expected for a phase that breaks a continuous rotational symmetry~\cite{Oganesyan2001,Vishwanath2014}. 
We therefore argue that a system of interacting fermions with a Rashba-like dispersion offers a simple, generic route to realize a non-Fermi liquid phase.

In contrast to the ferromagnetic and nematic phases, in the SDW phase the coupling between the fermions and the  Goldstone modes vanishes at long wavelengths (as it does for the case of CDW order~\cite{Kirkpatrick2009,Sun2009,Sun2009E,Vishwanath2014}). It is likely that it such coupling leads to qualitatively weaker effects compared to the ferromagnetic or nematic cases.

The nematic and spin-density wave phases may become superconducting at sufficiently low temperature; such an instability has been argued to be strongly enhanced in the presence of gapless nematic fluctuations.~\cite{Metlitski2014,Lederer2014} The ferromagnetic phase, however, does not posses a superconducting instability, since it breaks the symmetries of time reversal, inversion, and rotation by $\pi$ around the $z$ axis. Therefore, the non-Fermi liquid phase may be robust down to arbitrarily low temperatures.

It is interesting to comment on the effect of disorder on the different symmetry breaking phase. Non-magnetic disorder couples linearly to the nematic order parameter. The system therefore maps onto a random field XY model; therefore, the nematic phase is expected to be destroyed by disorder by the Imry-Ma argument.~\cite{Imry1975} However, since non-magnetic disorder does not couple linearly to the magnetic component of the order parameter, there is a possibility that the SDW and FM phases still posses a finite temperature transition in the presence of disorder. In the case of the ferromagnet, the system maps into the random anisotropy XY model.~\cite{Harris1973} It is an interesting open question whether an Ising finite-temperature transition can occur in this system at $d=2$ for weak disorder.

\acknowledgments{We would like to thank Ehud Altman, Gareth Conduit, Nir Davidson, Sarang Gopalakrishnan, Anna Kesslman, Daniel Podolsky, Jonathan Schattner, and Senthil Todadri for helpful discussions.
E. B. was supported by the Israel Science Foundation, by the Minerva foundation, and by an Alon fellowship. E. B. also thanks the Aspen Center for Physics, where part of this work was done. J. R. was supported by the ERC synergy grant UQUAM. }

{\it Note added.--} A related paper, Ref.~\onlinecite{Bahri2014}, has appeared in parallel to this work.  Our results are consistent where they overlap. 

\appendix

\section{Variational calculation}\label{app:var}

\begin{figure}
\centering
\includegraphics[width=9.cm,height=7.cm]{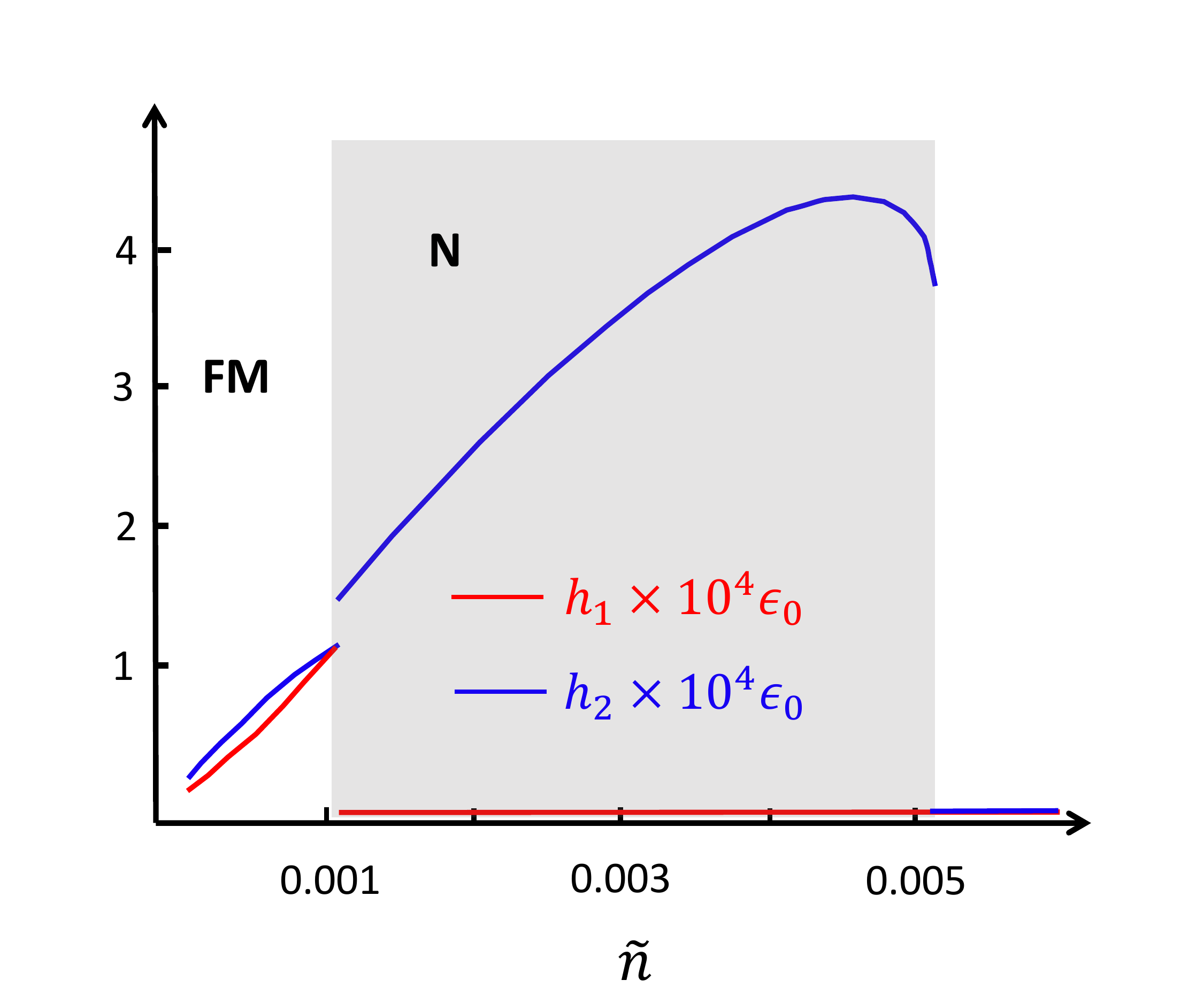}
\caption{ The mean-field variational parameters $h_1$ and $h_2$ vs. the density as obtained by
equations (\ref{VE1}-\ref{VE3}) for $\tilde U = 0.5$. }\label{fig:OP}
\end{figure}

In this appendix we derive the self-consistency equations (\ref{VE1}-\ref{VE3}) using the variational principle. We seek the best
candidate ground state for the Rashba Hamiltonian (\ref{H_0}) with the interaction term (\ref{V_Fourier})
\be
\mc H_I = {1\over \Omega^2} \sum_{\bs k,\bs k'}\sum_{l=0}^{\infty} V_l\,e^{i\phi_{\bs k \bs k'}}n_{\bs k} n_{\bs k'} \label{app:HI}
\ee
The variational trial state, $| \{ h_l \} \rangle $, is taken to be the ground state of the mean-field Hamiltonian
(\ref{MF_nematic}). The value of the variational parameters $h_l$ and $\mu$ are determined by minimizing the energy functional
$
E_V \equiv \langle\{h_l\}|\mc H_0 +\mc H_I |\{h_l\}\rangle
$
\[ {\partial E_V \over \partial h_l} = 0 \;\; \mrm{and}\;\;n = {1\over \Omega} \sum_{\bs k} \langle n_{\bs k} \rangle\]
To simplify the variational equations we use the identity ${\partial \over \partial h_l}\langle \mc H_{MF}^N \rangle = -n_l   $,
where $n_l \equiv    {1\over \Omega } \sum_{\bs k}\cos l \phi_{\bs k}\, \langle n_{\bs k} \rangle$
\begin{align}
{\partial E_V \over \partial h_l} = &{\partial \over \partial h_l}\left[ \langle \mc H_{MF}^N \rangle + \sum_{l'}\left( h_{l'} n_{l'}
+ V_{l'} n_{l'}^2\right)\right]\label{dhl}\\
=& \sum_{l'} \left( h_{l'} +2 V_{l'} n_{l'} \right) {\partial n_{l'} \over \partial h_l} = 0
\end{align}
Thus, we find that the minimum solution for $l=0,1,2$ is obtained by the equations (\ref{VE1}-\ref{VE3}) as long as the matrix
$Q_{ll'} = {\partial n_{l'} \over \partial h_l}$ is not singular.
The solution of the equations for $\tilde U = 0.5$ is presented in Fig.~\ref{fig:OP}.

We note that it is straightforward to generalize the derivation of the self-consistency equations (\ref{VE1}-\ref{VE3}) to the case of a textured order parameter (not translationally invariant). We simply substitute $n_l$ by
\be
 n_l (\bs q) =  \sum_{\bs k} \cos l \phi_{\bs k} \langle\psi_{\bs k + {\bs q \over 2}}^\dag \psi_{\bs k - {\bs q \over 2}} \rangle
\ee
in Eq.~(\ref{app:HI}) - Eq.~(\ref{dhl}). An important outcome of this generalization is that the coupling constant that couples to the SDW order $n_1(\bs q)$ is $V_1$. This validates using $V_1$ as the corresponding coupling constant in the Stoner criterion for the SDW instability in accord with the main text.

It is also useful to point out that the solution of the equations (\ref{VE1} - \ref{VE3}) can be simplified to some extent in the
case of a single order parameter. Performing the integration over $\bs k$ in the case of $h_2 =0$ we find that these equations reduce
to
\begin{align}
&n = {k_0 k_F ^x \over \pi^2  }\, E \left(  {2\over 1+ a}; {\phi_0 \over 2} \right)\\
&h_1 = {4 k_0 k_F ^x V_1 \over 3 \pi^2}\left[ a\,E\left(  {2\over 1+ a}; {\phi_0 \over 2} \right)+(1-a)\,K\left(  {2\over 1+ a};
{\phi_0 \over 2} \right) \right]
\end{align}
where $k_F ^x = k_0 \sqrt{h_1 + \mu \over \e_0}$, $a \equiv \mu / h_1$, $\phi_0 \equiv \arccos (-a)$ and $
E(x;\phi_0)=\int_0^{\phi_0} d\phi \sqrt{1-x^2 \sin^2 \phi}$ and $K(x;\phi_0) = \int_0^{\phi_0} d\phi {1\over\sqrt{1-x^2 \cos^2
\phi}}$ the partial elliptic integral of the second kind. Similarly, in the nematic case where $h_1=0$ we have
\be
h_2 = {2 k_0 k_F ^x V_2 \over 3 \pi^2}\left[ a\,E\left(  {2\over 1+ a}; {\phi_0 } \right)+(1-a)\,K\left(  {2\over 1+ a}; {\phi_0 }
\right) \right]
\ee
where $a = \mu/h_2$ and there are two Fermi surfaces.

\section{The spin susceptibility of the Rashba gas in the dilute limit}\label{app:spin susceptability}
In this appendix we calculate the in-plane spin-susceptibility of the Rashba gas, which is given by
\be
\chi_{ij}(i\w,\bs q)={1\over \Omega}\sum_{\bs k} {n_F (\xi_{\bs k}) - n_F (\xi_{\bs {k+q}})\over-i\w+ \e_{\bs {k+q}}-\e_{\bs k}}\,\mc
P(\phi_{\bs k} + \phi_{\bs {k+q}}) \label{chi}
\ee
where
\[\mc P(x) = {1\over 2}\begin{pmatrix} 1-\cos x & \sin x \\ \sin x & 1+ \cos x \end{pmatrix} \]

\begin{figure}
\centering
\includegraphics[width=6.5cm,height=5.5cm]{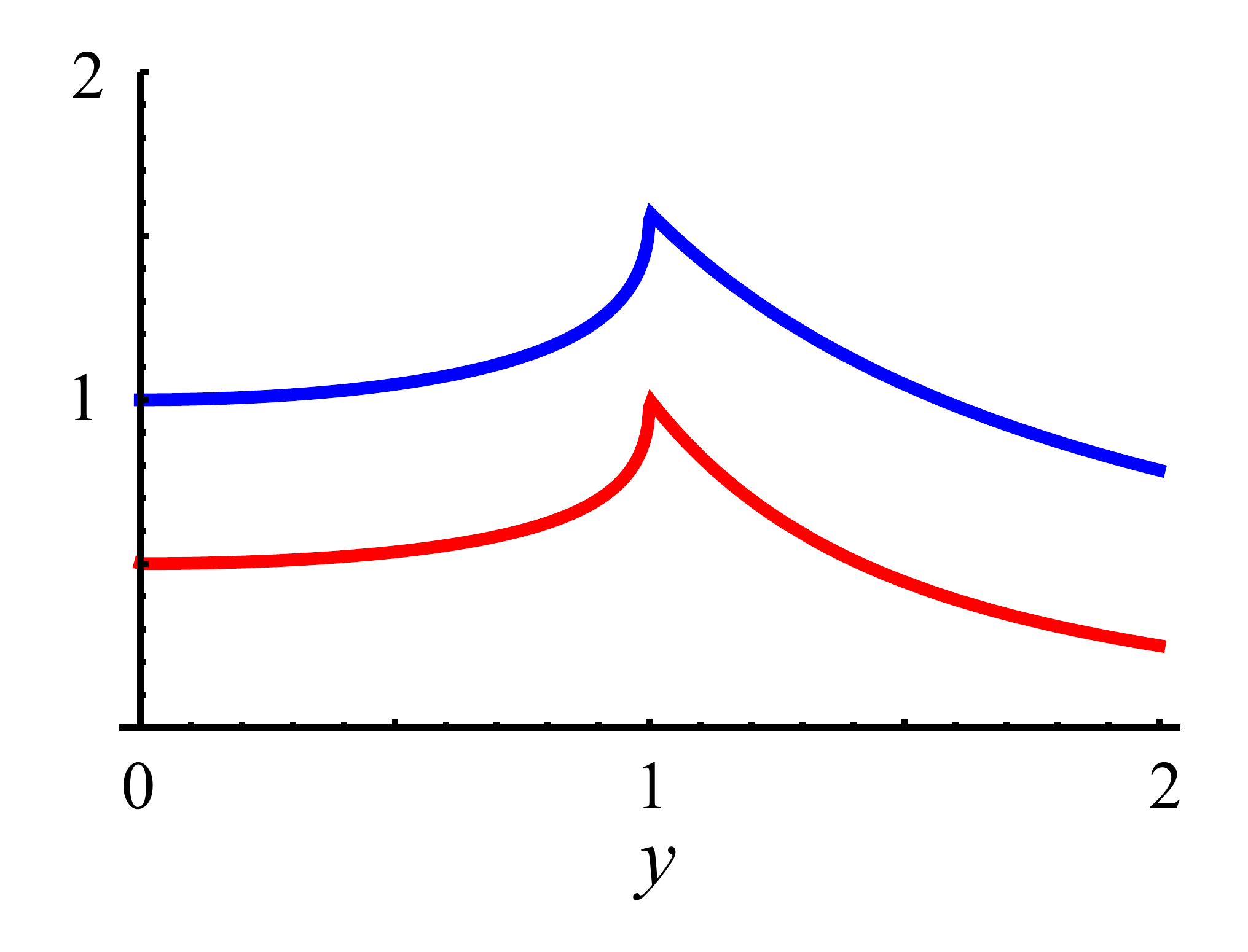}
\caption{ The functions $\mc F_{1}(0,y)$ (blue) and $\mc F_{-1}(0,y)$ (red) as a function of $y$. }\label{fig:F}
\end{figure}

In the FL phase we can compute (\ref{chi}) analytically in the static limit.
First, we linearize the denominator term $\e_{\bs {k+q}}-\e_{\bs k}\approx v_0 {q \d k \over k_0}  \cos \phi_{\bs k ,\bs q}  $ where
$v_0 = \e_0 / k_0$, $\d k = k - k_c$ and $ k_c \approx k_0\left(1-{q\over 2 k_0} \cos\phi_{\bs k, \bs q} \right)$. Integrating over
$\d k$ and $\phi_{\bs k}$ yields
\begin{widetext}
\be
\chi_{ij}(i\w,\bs q)={\rho\left(\mu \right)\over 2} \,
\left( \mc F_{-1} \left(\tilde \w,\tilde q\right) -\cos {2\phi_{\bs q} }\left[ 2\mc F_{1} \left(\tilde \w,\tilde q\right)-\mc F_{-1}
\left(\tilde \w,\tilde q\right) \right]\s^z\right)
\ee
where $\tilde \w \equiv {m\w \over 2k_F q }$, $\tilde q \equiv {q\over 2k_F}$ and
\[\mc F_n\left(x,y \right) =  {1\over 8\pi}\int_{0} ^{2\pi} {d\phi} \,cos^n \phi \,\log\left({x^2 + (1+y \cos \phi)^2 \cos^2 \phi
\over x^2 + (1-y \cos \phi)^2 \cos^2 \phi} \right) \]
\end{widetext}
\[\]
\[\]
In the static limit $\w \rightarrow 0$ we obtain
\begin{align}
&\mc F_{1}(0,y) = {1\over y^2} \bigg\{ \begin{matrix} 1-\sqrt{1-y^2} & ;& y<1 \\ 1&; & y>1 \end{matrix} \nn\\
&\mc F_{-1}(0,y) = {1\over y} \bigg\{ \begin{matrix} \arcsin y &\;\;\;
\;\; &; &  y<1 \\ \pi /2 &\;\;\;\, &; & y>1 \end{matrix}\,. \nn
\end{align}
These functions are both sharply peaked at $y = 1$ (see Fig.~\ref{fig:F}).

\end{document}